\documentclass[a4paper] {article}
\usepackage{graphics,color,epsfig}

\def\linadj#1{\normalbaselines
	\multiply\lineskip#1 \divide\lineskip100
 	\multiply\baselineskip#1 \divide\baselineskip100
	\multiply\lineskiplimit#1 \divide\lineskiplimit100 }
\newcommand{\n}{\noindent}

\begin{document}

\title{\bf Thermal and interfacial properties of a Quark Gluon Plasma droplet in a hadronic medium via a statistical model}

\author{ R. Ramanathan$^1$, Agam K. Jha$^1$, S. S. Singh$^1$ and K. K. Gupta$^+$}
\maketitle

\begin{center}

 \large $^1$ Department of Physics, University of Delhi, Delhi - 110007, INDIA.
\\  \large  $^+$Department of Physics, Ramjas College, University of Delhi, Delhi - 110007, INDIA.
\end{center}

\linadj{200}

\n{\bf  Abstract:}
\large  Thermal and interfacial properties of a QGP droplet in a hadronic medium  are computed using a statistical model of the system. The results indicate a weakly first order transition at a transition temperature $\sim (160 \pm 5)~MeV$. The interfacial surface tension is proportional to the cube of the transition temperature irrespective of the the magnitude of the transition temperature. The  velocity of sound in the QGP droplet is predicted to be in the range $(0.27~\pm~ 0.02)$ times the velocity of light in vacuum, and this value is seen to be independent of the value of the transition temperature as well as the model parameters. These predictions are in remarkable agreement with Lattice Simulation results and extant MIT Bag model predictions.\\

PACS number(s): 25.75.Ld, 12.38.Mh, 21.65.+f   
\vfill
\eject

\section {Introduction}
\label {sect1} 

\large  The formation of QGP droplet (fireball) is one of the most exciting possibility in the ultra relativistic heavy ion collision (URHIC) [1]. The physics of such an event is very complicated and to extract meaningful results from a rigorous use of QCD appropriate to this physical system is almost intractable though heroic efforts at lattice estimation of the problem has been going on for quite some time [2]. One way out is to replicate the approximation schemes which have served as theoretical tools in understanding equally complicated atomic and nuclear systems in atomic and nuclear physics in the context of QGP droplet formation. This approach lays no claim to rigour or ab-initio ``understanding'' of the phenomenon but lays the framework on which more rigorous structures may be built depending on the phenomenological success of the model as and when testable data emerges from ongoing experiments.

Our intention in this paper is to reevaluate the free energy of a QGP droplet in a bulk hadronic (pionic) medium, 
again in the limit of vanishing chemical potential, but using a different semi-phenomenological model for the system. 
The MIT bag model is simplicity itself; it puts all quarks and gluons as free particles inside a bag and makes the 
impermeable bag as the agent of confinement by ascribing a set of boundary conditions for quarks and gluons. It is 
fine to use the MIT bag model to describe the hadrons as bags of quarks, antiquarks and gluons, but to extend 
the idea to represent the phase boundary between the QGP droplet and the bulk hadronic medium makes one a bit uneasy, though it is plausible.
And, this is precisely the assumption made by the earlier authors [3] who have used the MIT bag model for the system 
of QGP droplet in a bulk hadronic medium.It is our intension to seek an alternative approach to the MIT bag model which does away with the assumption i.e; the confining bag 
of the hadrons has the same property as the interface separating the two phases, we propose an alternative model to represent the same physical situation. 
\section {The template density of states of Thomas-Fermi and Bethe and their QGP variant} 
\label {sec2}
In a very elegant and successful statistical model of atoms of large atomic numbers Thomas and Fermi [7] demonstrated 
the way to compute electronic density of states to very high order of accuracy. The Thomas Fermi model of atom assumes 
the electrons to be Fermi-Dirac gas confined within a localized region by the confining electrostatic potential V(r) 
of the central nucleus. The potential is assumed to be very slowly varying in the region with the average thermal 
energy  T (setting the Boltzmann constant to unity) is small compared to V(r) within  the region and comparable to 
it near boundary.

It is  now straight forward [7] to compare the electronic density of states, assuming all states to be filled in a 
volume $\nu$

\begin{equation}
N_e = p_{max}^3 ~  \nu / 3 \pi^2
\end{equation}

 The maximum kinetic energy of the electron at any point in phase space should not exceed the electrostatic potential 
(confining) at that point and therefore \newline $p_{max}^2 / 2m = - V(k)$, when $k$ is the phase point under consideration and $V(k)$ is the momentum transform of the coordinate potential $V(r)$. Therefore, the total density of states in phase space is given by

\begin{equation}
\int{\rho_{e}}(k)dk =  [ -2m V(k)]^{3/2} \nu / 3 \pi^2
\end{equation}
 or,
\begin{equation}
\rho_e (k) = [\nu (2m)^{3/2} / 2 \pi^2]~ [-V(k)]^{1/2} \cdot \biggl[-\frac{dV(k)}{dk}\biggl]
\end{equation}

In a modified statistical model, density of states for nucleons Bethe[7]gave a formula with a chosen potential as

\begin{equation}
\rho_{A}(k)=\frac{\pi^{1/2}}{12k^{5/4}a^{1/4}}exp(2\sqrt {ak})
\end{equation}
where `a' is the density paprameter which satisfies
\begin{equation}
 \rho_{Fermi-Surface}=\frac{1}{6}\pi^{2}a 
\end{equation}
for single particle level density.

 We can adapt the basic ideas of these models to the case of a QGP droplet, the electrons or nucleons get replaced by quarks which 
are also Fermions, and the minimum kinetic energy of the quarks at each point in phase space must exceed the confining/
de-confining potential at that point, since the QGP by definition is a deconfined gas  of relativistic quarks and gluons 
as against the non-relativistic electron of the conventional Thomas-Fermi Model or the nucleon of the bethe model. Therefore, $p_{min}=[-V_{conf}(k)]$ 
and $p_{max}=[-V_{conf}(\infty)]$ which represents a reference energy and can be set to zero, remembering that we 
are dealing with a relativistic system and where $`k'$ refers to the corresponding quark momenta in phase-space. So an expression similar to (3) holds for the  quark density of states, with the replacement of $V(k)$ with a suitable QCD induced phenomenological potential. The quark density  of states 
therefore is 

\begin{equation}
\rho_q (k) = (\nu / \pi^2) [-V_{conf}(k)]^2 \biggl[\frac{dV_{conf}(k)}{dk}\biggl]
\end{equation}

 In this adaptation of the Thomas-Fermi and the Bethe ideas, we only capture the spirit 
of the original idea for a system which is very different in detail.

It is easy to observe that this is a typical first order phase transition behaviour indicated by the Ramanathan et.al approach, thereby allowing direct comparision and the opportunity to use the calculational advantages of each of the approaches to bolster the usability of the approximation schemes as a phenomenological tool in the analysis of the data as we shall see in the following.

In the approximation schemes of the Ramanathan et. al [5], the relativistic density of states for the quarks and gluons is constructed adapting the procedures of the Thomas-Fermi construction of the electronic density of states for complex atoms and the Bethe density of states [7] for nucleons in the complex nuclei as templates. The expression for the density of states for the quarks and gluons (q, g)in this model is   

\begin{equation}
\rho_{q,g}(k) = (v / \pi^2)\biggl[(-V_{conf}(k))^2 \biggl(\frac{dV_{conf}(k)}{dk}\biggl)\biggl]_{q,g}
\end{equation}  

where $k$ is the relativistic four-momentum of the quarks and gluons, $v$ is the volume of the fireball taken to be a constant in the first approximation and$V_{conf}$ is a suitable confining potential relevant to the current quarks and gluons in the QGP [5] given by

\begin{equation}
[V_{\mbox{conf}}(k)]_{q,g} = (1/2k)\gamma_{q,g} ~ g^2 (k) T^2 - m_0^2 / 2k
\end{equation}

where $m_{0}$ is the mass of the quark which we take as zero for the up and down quarks and $150 MeV$ for the strange quarks. The $g(k)$ is the QCD running coupling constant given by 

\begin{equation}
g^2(k) = (4/3) (12\pi/27) (1/ ln(1+k^2/\Lambda^2))
\end{equation}                                                                       

where $\Lambda $ is the QCD scale taken to be $150~ MeV$.

The model has a natural low energy cut-off at 

\begin{equation}
  k_{min} = (\gamma_{q,g} N^{1/3} T^2 \Lambda^{2}/2)^{1/4}
\end{equation}                                                                                                                               
with $$N = (4/3)(12\pi/27).$$

The above cut off is equivalent to a deconfinement condition wherein the energy of the quark and gluon at each point in phase space has to exceed the value of the confining potential at that point.
  
  The freee energy of the respective case $i$ (quarks, gluons, interface etc.) for Fermions and Bosons (upper sign or lower sign)can be computed using the following expression

 \begin{equation}
F_i = \mp T g_i \int dk \rho_i (k) \ln (1 \pm e^{-(\sqrt{m_{i}^2 + k^2}) /T})
\end{equation}                                                                       
with the surface free-energy given by a modified Weyl [8] expression:

\begin{equation}
F_{surface} = \frac{1}{4}  R^{2}  T^{3} \gamma
\end{equation}                                                                       
where the hydrodynamical flow parameter for the surface is :

\begin{equation}
\gamma = \sqrt{2}\times \sqrt{(1/\gamma_g)^2+(1 / \gamma_q)^2},
\end{equation}                                                                    
For the pion which for simplicity represents the hadronic medium in which the fireball resides, the free energy is :

 \begin{equation}
F_{\pi} = (3 ~ T/2\pi^2 )(-v) \int_0^{\infty} k^2 dk \ln (1 - e^{-\sqrt{m_{\pi}^2 + k^2} / T})
\end{equation}

With these ingredients we can compute the free-energy change with respect to both the droplet radius and temperature to get a physical picture of the fireball formation, the nucleation rate governing the droplet formation, the nature of the phase transition etc. This can be done over a whole range of flow-parameter values [5], We exhibit only the two most promising scenarios in fig.$1$ and fig.$2$.
\begin{figure}
\epsfig{figure=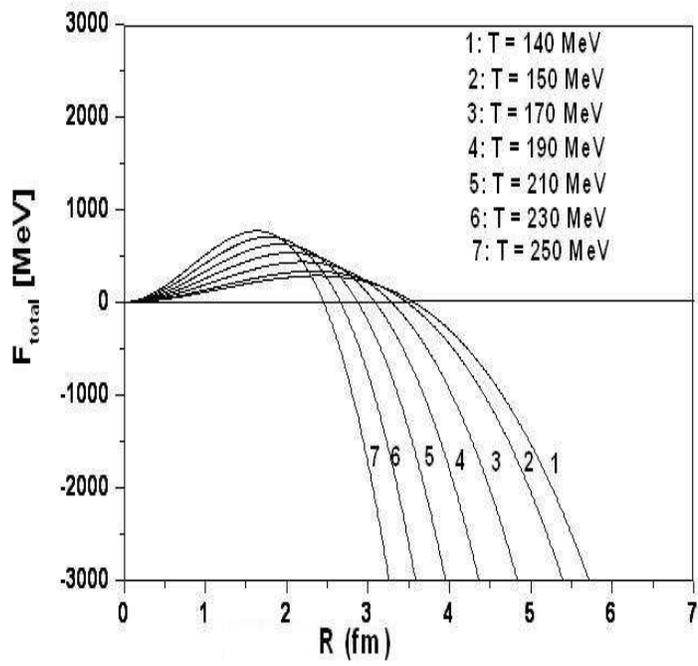,height=4.5in,width=4.0in}
\label{fig5}
\caption{\large $F_{total}$ at $\gamma_{g} = 8\gamma_{q}$, $ \gamma_{q} = 1/6 $ for various temperatures.}
\end{figure}

\begin{figure}
\epsfig{figure=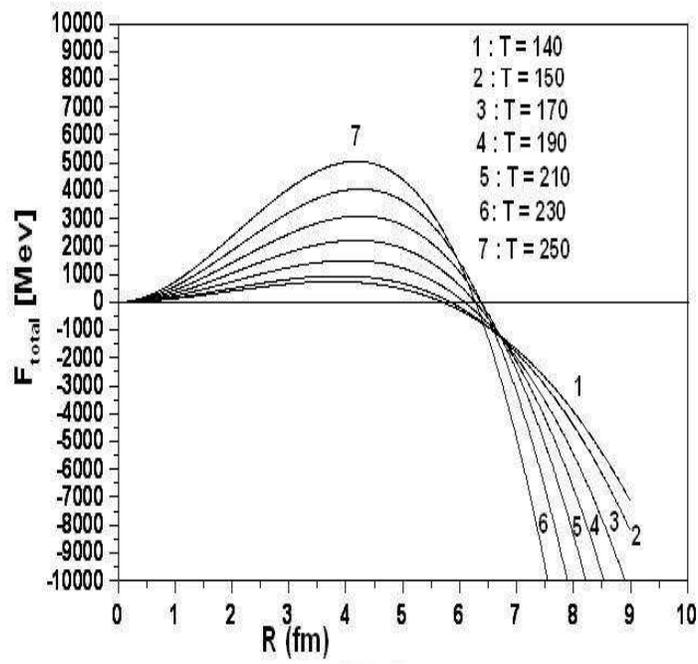,height=4.5in,width=4.0in}
\label{fig4}
\caption{\large $F_{total}$ at $\gamma_{g} = 6\gamma_{q}$ , $ \gamma_{q} = 1/6 $ for various temperatures.}
\end{figure}
\vfill
\eject

\section {Thermodynamical behaviour of the droplet and the nature of phase transition}
\label {sec3}
Standard thermodynamics gives the following relations:
\begin{equation}
 Entropy ~  S =-({\partial{F}}/{\partial{T}})
\end{equation}

\begin{equation}
 Specific ~ heat ~at~ constant~ volume  ~  C_{V} =T( {\partial{S}}/{\partial{T}})_{V}
\end{equation}

\begin{equation}
Sound-Velocity ~    C_{S}^{2}={S}/{C_{V}}
\end{equation} 

After plugging the total free energy in the previous section into these expressions, we evaluate these quantities for the two promising scenarios of our model, the result of which are displayed in the following figs. 3 to 8.  
 
\begin{figure}
\epsfig{figure=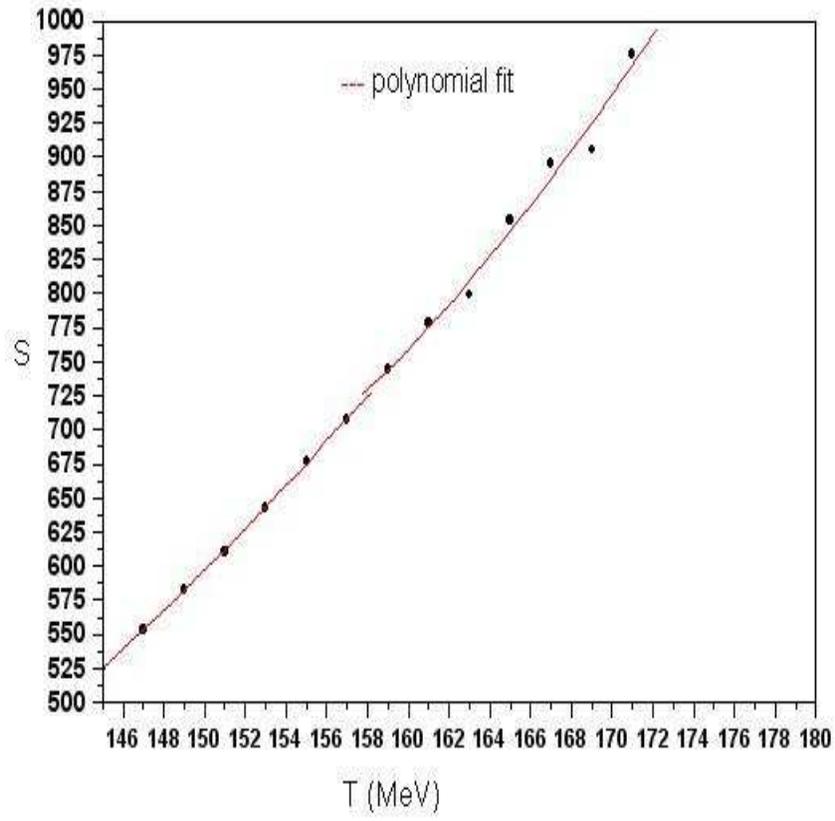,height=5.5in,width=5.0in}
\label{st61.eps}
\caption{\large  Variation of $S$ with temperature T at $\gamma_{g} = 6\gamma_{q}$ , $ \gamma_{q} = 1/6 $.}
\end{figure}
 \begin{figure}
\epsfig{figure=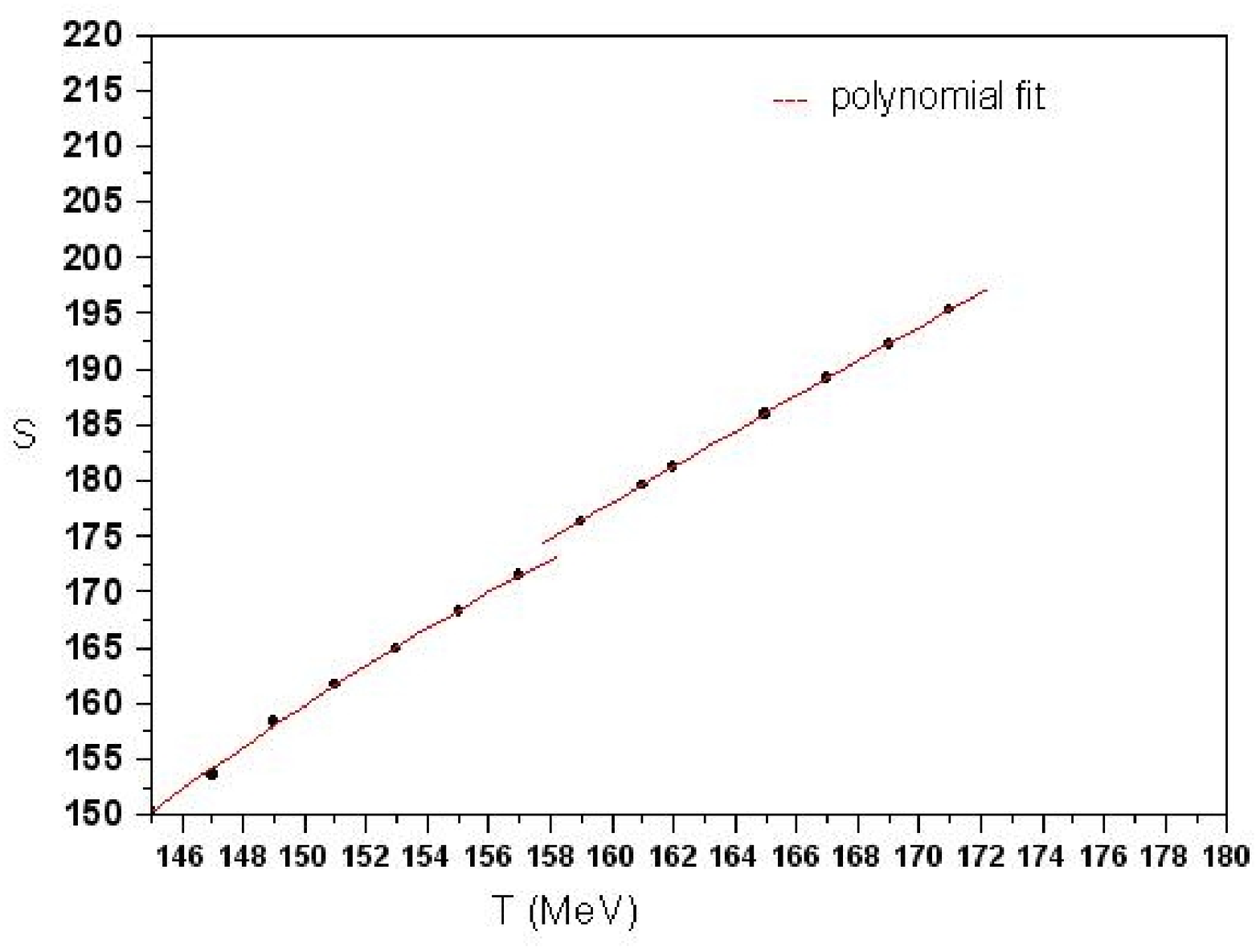,height=5.0in,width=4.5in}
\label{st81.eps}
\caption{\large  Variation of $S$ with temperature T at $\gamma_{g} = 8\gamma_{q}$ , $ \gamma_{q} = 1/6 $.}
\end{figure}
 \begin{figure}
\epsfig{figure=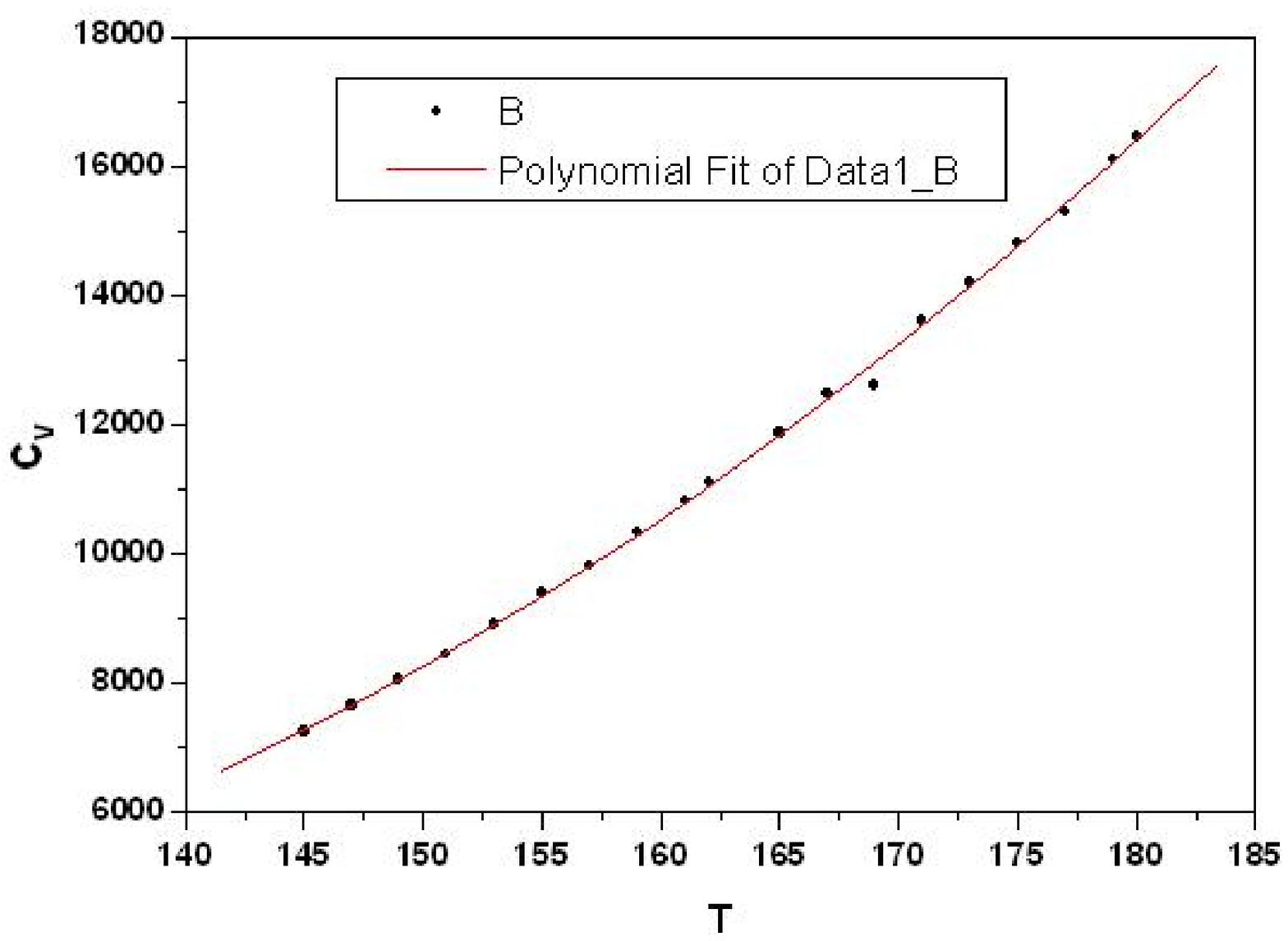,height=4.5in,width=4.0in}
\label{cv6.eps}
\caption{\large Variation of specific heat $C_{V}$ with temperature T at $\gamma_{g} = 6\gamma_{q}$ , $ \gamma_{q} = 1/6 $.}
\end{figure}

\begin{figure}
\epsfig{figure=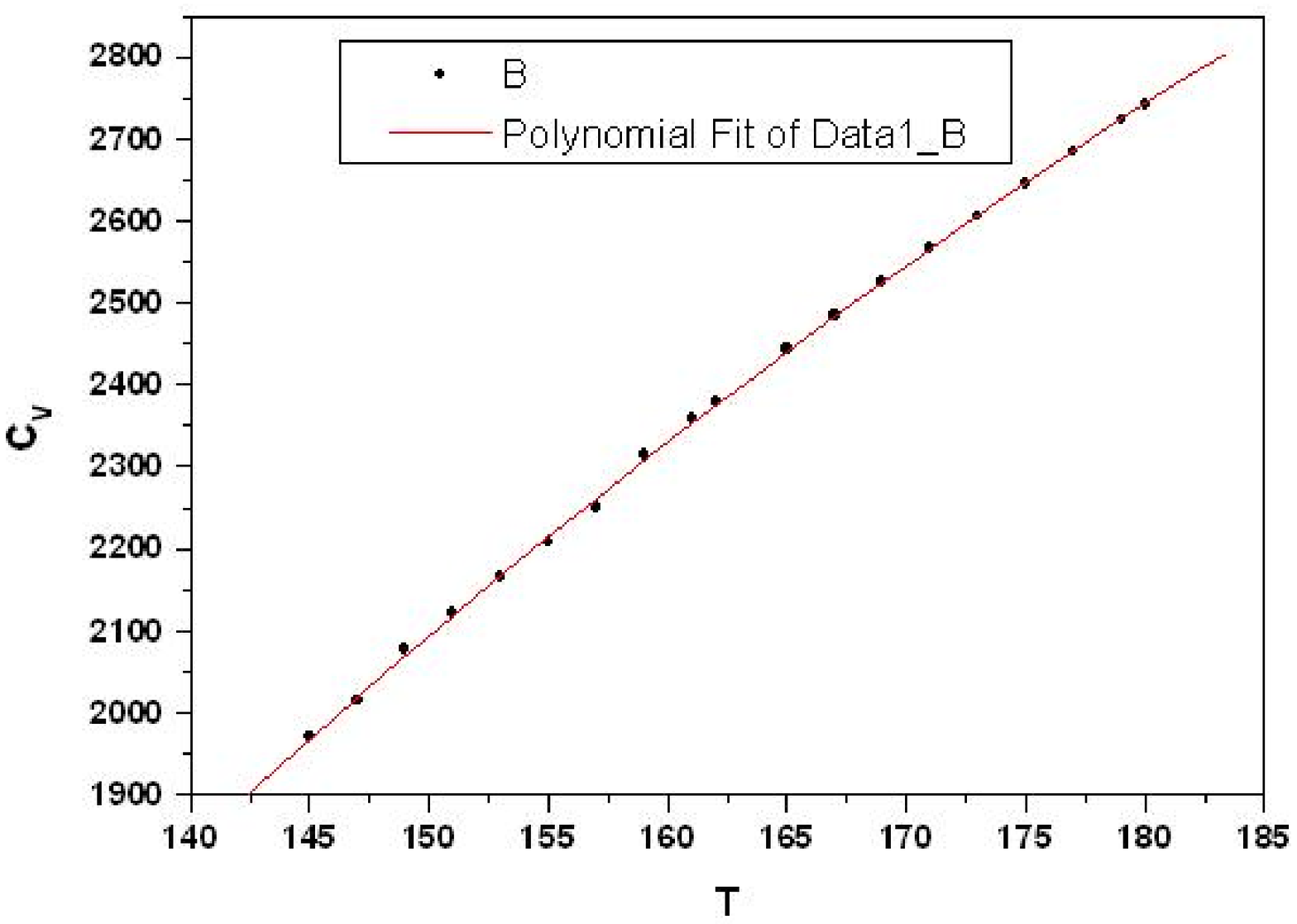,height=5.0in,width=4.5in}
\label{cv8.eps}
\caption{\large Variation of specific heat $C_{V}$ with temperature T at $\gamma_{g} = 8\gamma_{q}$ , $ \gamma_{q} = 1/6 $.}
\end{figure}

\begin{figure}
\epsfig{figure=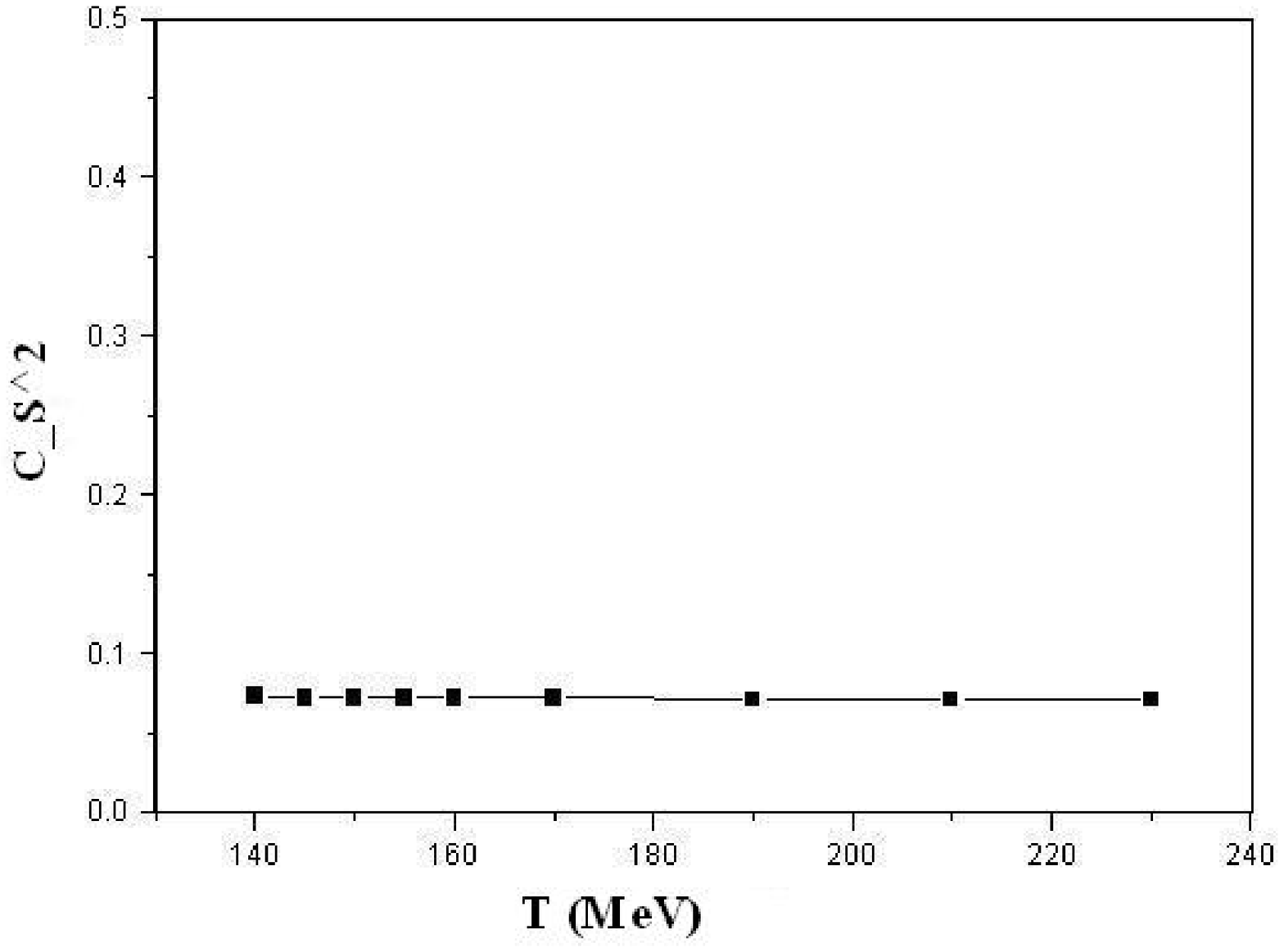,height=5.0in,width=4.5in}
\label{v6.eps}
\caption{\large Variation of velocity of sound squared $ C_{S}^{2}$ with temperature T at $\gamma_{g} = 6\gamma_{q}$ , $ \gamma_{q} = 1/6 $.}
\end{figure}
\begin{figure}
\epsfig{figure=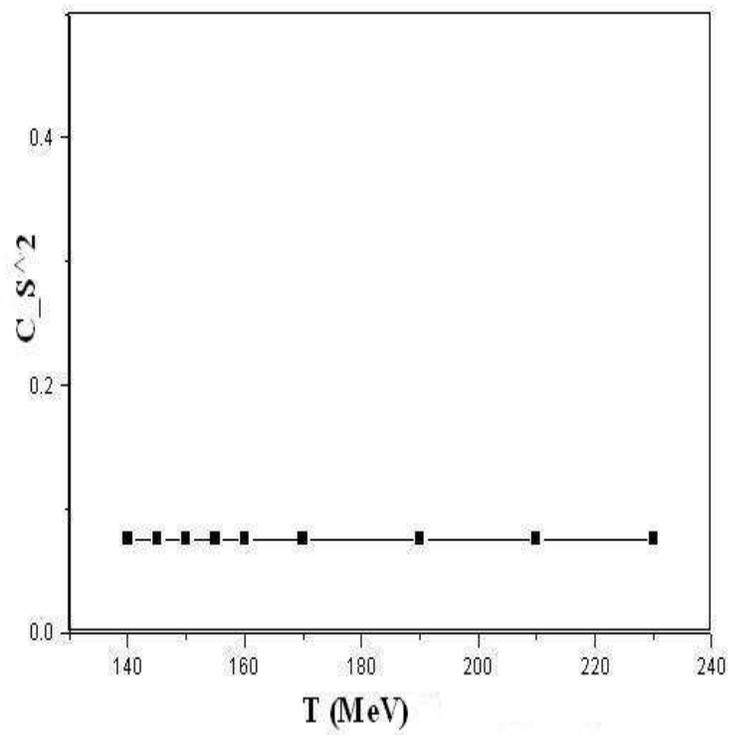,height=5.0in,width=4.5in}
\label{v8.eps}
\caption{\large Variation of velocity of sound squared $C_{S}^{2}$ with temperature T at $\gamma_{g} = 8\gamma_{q}$ , $ \gamma_{q} = 1/6 $.}
\end{figure}
\vspace{5.0cm}
\vspace{8.0cm}
\section {Surface tension of the interface}
\label {sec4}
 
The nucleation process is driven by statistical fluctuations being determined by the critical free energy difference between two phases. The Kapusta et. al model [4] uses the liquid drop model expansion for this , as given by

\begin{eqnarray} 
\Delta F = \frac {4\pi}{3} R^{3} [P_ {had}(T,\mu_{B}) - P_{q,g}(T, \mu_{B})] \nonumber \\
  + 4\pi R^{2} \sigma +\tau_{crit} T~ln \biggl [ 1 + (\frac {4 \pi}{3})R^{3}s_{q,g} \biggl]                                                                     \end{eqnarray}   

 The first term represents the volume contribution, the second term is the surface contribution where $\sigma$ is the surface tension, and the last term is the so called shape contribution . The shape contribution is an entropy term on account of fluctuations in droplet shape which we may ignore in the lowest order approximation. The critical radius $ R_{c}$ can be obtained by minimising (18) with respect to the droplet radius $R$ , which in the Linde approximation [4] is,

 \begin{equation}
 R_{c} = \frac {2\sigma}{\Delta p}~ or~                                             \sigma = \frac {3\Delta F_{c}}{4\pi R_{c}^{2}}
\end{equation}
\begin{quote}
\begin{tabular}{|r|r|r|r|r|}
\hline
$T_{c}$&$\Delta F_{c}$&$R_{c}$&$\sigma$&$\frac{\sigma}{T_{c}^{3}}$\\  
$(MeV)$&$(MeV)$&$(fm)$&$(MeV/fm^{2})$&\\
\hline
150&332.203&3.475&6.568&0.078\\
160&382.359&3.385&7.966&0.078\\
170&433.037&3.285&9.580&0.078\\
190&532.219&3.085&13.35&0.078\\
210&623.349&2.875&18.004&0.078\\
230&702.041&2.655&23.776&0.078\\
250&766.041&2.455&30.343&0.078\\
\hline
\end{tabular}
\end{quote}
Table-1 for Surface Tension of QGP droplet at $\gamma_{g}=8\gamma_{q}$,$ \gamma_{q} = 1/6$.
\vfill
\begin{quote}
\begin{tabular}{|r|r|r|r|r|}
\hline
$T_{c}$&$\Delta F_{c}$&$R_{c}$&$\sigma$&$\frac{\sigma}{T_{c}^{3}}$\\
$(MeV)$&$(MeV)$&$(fm)$&$(MeV/fm^{2})$&\\
\hline
150&943.595&5.835&6.616&0.078\\
160&1197&5.965&8.031&0.078\\
170&1494&6.085&9.633&0.078\\
190&2216&6.275&13.435&0.078\\                                               
210&3088&6.375&18.140&0.078\\                                                 
230&4059&6.375&23.844&0.078\\
250&5052&6.275&30.630&0.078\\
\hline
\end{tabular}                                       
\end{quote}
Table-2 for Surface Tension of QGP droplet at $\gamma_{g}=6\gamma_{q}$,$ \gamma_{q} = 1/6$.

The tables $1$ and $2$ list the surface tension computed in the Ramanathan et.al model which can be used in the dynamical models[4], thus reducing the arbitrariness in the latter models to this extent, thus enabling us to use the parameter extracted from a static situation to make perdictions about the dynamical growth process of fireball formation.      
\section {QGP droplet formation rate}
\label {sec5}
In this section we discuss the strength and weakness of two approximation schemes which seem promising in their usefulness in the above context and also explore their parametric interrelations which may be useful in phenomenological applications.The two approximation schemes under consideration are the QCD oriented QGP- droplet formation model of Csernai and Kapusta and its later refinements [4] and the simple statistical model of Ramanathan et. al [5] which again is essentially  QCD oriented in that use is made of an effective QCD potential. In both these schemes it is possible to estimate the fireball radius, the transition temperatures for their formation, nucleation rate etc. in quantitative terms for comparision with raw data as and when it is available from ongoing URHIC experiments.

As we shall see the two approaches can complement each other in the event of analysing the data on fireball formation especially with regard to nucleation of QGP droplets in a hadronic medium.

Central to this comparision is the rate $I$ to nucleate droplets of QGP in a hadronic gas per unit time per unit volume is given by [6] 

\begin{equation}
 I=I_{0}~e^{-\Delta F_{c}/T_{c}}
\end{equation} 

Where $I_{0}$ is the nucleation rate at vanishing change in free energy $ \Delta{ F} $ of the system due to the formation of a single critical size droplet of plasma. In the Kapusta et. al [3] formulation it is not possible to estimate the whole prefactor in terms of other possible measurable parameters of the droplet. 
There is no way to estimate the value of the crucial input $\sigma$ within the Kapusta et. al approach and the related approaches, though a value of $50~MeV/fm^{2}$ is assumed. The critical free energy in this model varies as in fig.$9$ [4].
\begin{figure}
\epsfig{figure=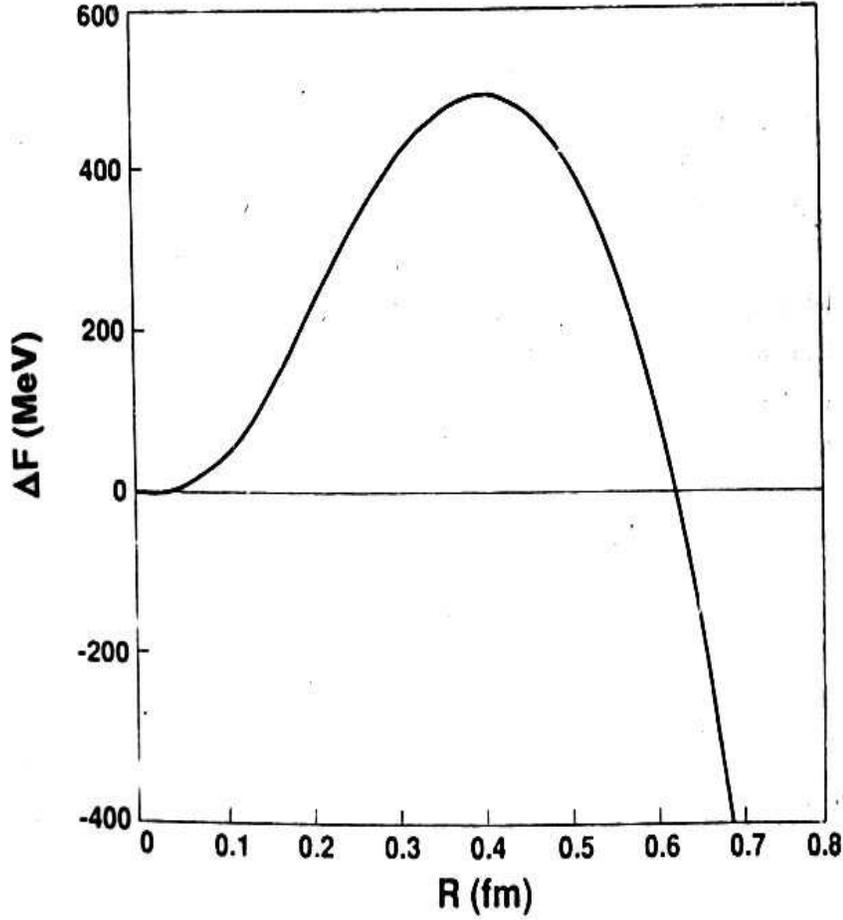,height=5.0in,width=4.5in}
\label{figure1}
\caption{\large The free energy difference $\Delta F(R)$ between a hadronic phase with and without a quark-gluon plasma droplet [4]}
\end{figure}

 As could be seen the general nature of the curves given by figs.$1$, $2$ and $9$, though using two totally different approches leads to the conclusion that both predict a first order phase transition for the fireball production process. As a derivative computation we can also compute the nucleation rate leading to the droplet formation, which for the two scenarios is illustrated by figs.$ 10$  and $11$.

\begin{figure}
\epsfig{figure=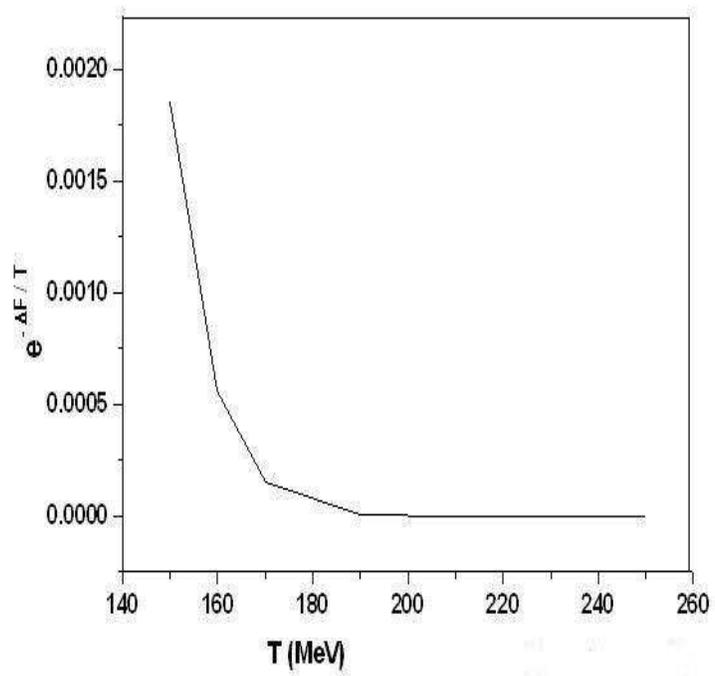,height=4.5in,width=4.0in}
\label{exp2}
\caption{\large The fireball formation rate with temperatures  at $\gamma_{g} = 6\gamma_{q}$, $ \gamma_{q} = 1/6 $ .}
\end{figure}

\begin{figure}
\epsfig{figure=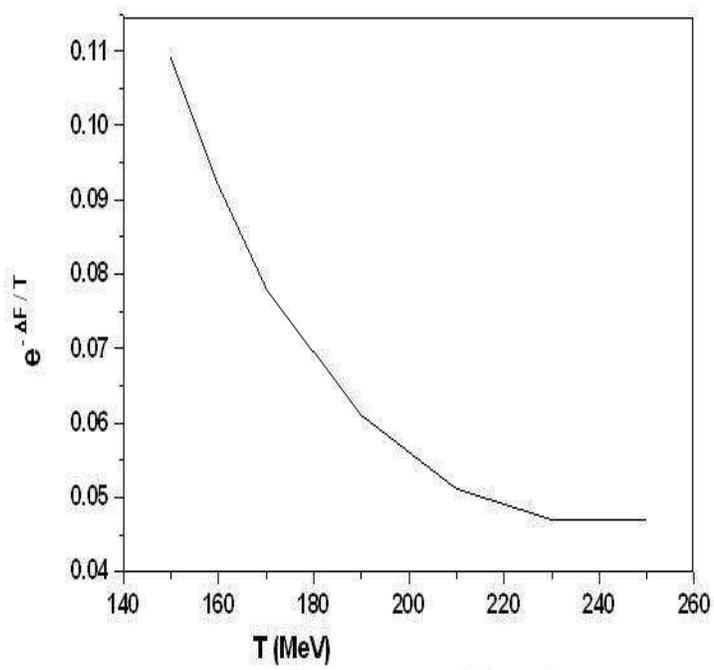,height=4.5in,width=4.0in}
\label{exp3}
\caption{\large The fireball formation rate with temperatures at $\gamma_{g} = 8\gamma_{q}$ , $ \gamma_{q} = 1/6 $ .}
\end{figure}

As could easily be observed from figs 10 and 11, for both the sets of parameter values, the droplet formation rate grows exponentially in the vicinity of transition temperature range of $150$ to $170~ Mev$ as expected from lattice simulations and other model calculations. The critical free-energy and radius at the maximum of the curves in figs. 1 and 2 allow us to compute the surface tension used in the Kapusta et.al and the derivative models [4] as already shown in section $4$.
\section {Conclusion}
\label {sec6}

The graphs (Figs. 3 to 6) clearly indicate that the model predicts a weakly first order transition at a temperature in the range $(160~\pm~5)~MeV$, which seems to be consisitent with current expectations of QGP-Hadron phase transition [1]. This feature is also corroborated by the behaviour of the nucleation rate exhibited in Figs $10$ and $11$, where the exponential rise in the droplet formation rate in the vicinity of $160~MeV$, is indicative of the occurence of phase transition at these tempeartures.

The independence of the velocity of sound in the QGP system,Figs $7$ and $8$, from both the values of the model flow parameters as well as the magnitude of the transition temperature is remarkable. The value of the velocity of sound is consistently predicted to be of the order of $(0.27\pm0.02)$ times the velocity of light in vacuum. 
                                                                                 The tables $1$ and $2$ list the surface tension computed in the Ramanathan et.al model which can be used in the dynamical models[4], thus reducing the arbitrariness in the latter models to this extent, thus enabling us to use the parameter extracted from a static situation to make perdictions about the dynamical growth process of fireball formation.      

In both tables $1$ and $2$ the surface tension is seen to increase with the temperature of the fireball,which is a beautiful demonstration of a QCD effect.As the temperature of the QGP droplet increases the shear forces on the fireball surface will also increase tending to tear the surface quarks apart, consequently, bringing into play the confining property of the QCD forces manifesting itself in increased surface tension, which is exactly what the calculations show.Another striking feature of the result is the independence of the QGP droplet surface tension $\sigma$ variations in the flow parameters of the model and it varies with only the temperature, in the lowest order approximation we have employed.

The constancy of the ratio $\frac {\sigma}{T_{c}^{3}}$ indicates a cubic crtitical temperature dependence of the surface tension of the interfacial separation between the two phases. This is in striking conformity with the results of Lattice QCD simulation [9].

Finally it is readily seen from Tables $1$ and $2$, that at a transition temperature of $156~MeV$, the surface tension is predicted to be of the order of $67~MeV/fm^{2}$, a value in toatl agreement with the MIT bag model prediction of Abbas et.al [10].     

{\bf References :}
\begin{enumerate}
\item{F. Karsch, E. Laermann, A. Peikert, Ch. Schmidt and S. Stickan, Nucl. Phys. B(Proc. Suppl.)94, 411 (2001)}.
\item{T. Renk, R. Schneider, and W. Weise, Phys. Rev. C66, 014902 (2002)}.
\item{G. Neerguard and J. Madsen, Phys. Rev. D, 60, 054011 (1999); R. Balian and C. Block, Ann. Phys. (N. Y.), 64, 401(1970).}
\item{L. P. Csernai, J. I. Kapusta, E. Osnes, Phys. Rev. D67, 045003 (2003), e-Print Archive: hep-th/0201024; J. I. Kapusta, R. Venugopalan, A. P. Vischer, Phys. Rev. C51, 901-910 (1995), e-Print Archive: nucl-th/9408029; P. Shukla and A. K. Mohanthy, Phys. Rev. C 64, 054910(2001)}.
\item{R. Ramanathan, Y. K. Mathur, K. K. Gupta, and Agam K. Jha, Phys. Rev. C 70, 027903 (2004); R. Ramanathan, Y. K. Mathur, K. K. Gupta and S. S. Singh, e-print Archive: hep-ph/0502046}.
\item{  B. D. Malhotra and R. Ramanathan, Phys. Lett. A 108, 153,1985}.
  \item{E. Fermi, Zeit F. Physik 48, 73 (1928); L. H. Thomas , Proc. camb. Phil. Soc. 23, 542 (1927);H. A. Bethe, Rev. Mod. Phys. 9, 69 (1937)}.
\item{H. Weyl, Nachr. Akad. Wiss Gottingen 110 (1911)}.
\item{Y. Iwasaki, K. Kanaya, Leo Karkkainen, K. Rummukainen, and T. Yoshie, Phys. Rev. D49, No. 7, (1994) 3540; M. Hasenbusch, K. Rummukainen and K. Pinn, e-print arXiv: hep-lat/9312078}.
\item{L. Paria, M. G. Mushtafa, A. Abbas, Int. Journ. of Mod. Phys. E9(2000)149.} 
\end{enumerate}

\end{document}